\documentclass[draft,tightenlines,nofootinbib,preprint,aps,eqsecnum,amsmath,amssymb]{revtex4}

\newcommand{\beq}{\begin{equation}}
\newcommand{\eeq}{\end{equation}}
\newcommand{\bea}{\begin{eqnarray}}
\newcommand{\eea}{\end{eqnarray}}

\newcommand{\sgn}{\epsilon}

\begin{document}

\title{Hamiltonian ADM Gravity in Non-Harmonic Gauges with Well Defined Non-Euclidean 3-Spaces:
How Much Darkness can be Explained as a Relativistic Inertial
Effect?}

\medskip

\author{Luca Lusanna}

\affiliation{ Sezione INFN di Firenze\\ Polo Scientifico\\ Via Sansone 1\\
50019 Sesto Fiorentino (FI), Italy\\ Phone: 0039-055-4572334\\
FAX: 0039-055-4572364\\ E-mail: lusanna@fi.infn.it}

\begin{abstract}

In special and general relativity the synchronization convention of
distant clocks may be simulated with a mathematical definition of
global non-inertial frames (the only ones existing in general
relativity due to the equivalence principle) with well-defined
instantaneous 3-spaces. For asymptotically Minkowskian Einstein
space-times this procedure can be used at the Hamiltonian level in
the York canonical basis, where it is possible for the first time to
disentangle tidal gravitational degrees of freedom from gauge
inertial ones. The most important inertial effect connected with
clock synchronization is the York time ${}^3K(\tau, \sigma^r)$, not
existing in Newton gravity. This fact opens the possibility to
describe some aspects of {\it darkness} as a relativistic inertial
effect in Einstein gravity by means of a Post-Minkowskian
reformulation of the Celestial Reference System ICRS.

\bigskip

Talk at the Workshop "Cosmology, the Quantum Vacuum and Zeta
Functions" for the celebration of Emilio Elizalde's sixtieth
birthday, Barcelona, March 8-10, 2010.

\today

\end{abstract}

\maketitle

\vfill\eject

In classical and quantum physics predictability is possible only if
the relevant  partial differential equations have a well-posed
Cauchy problem, whose pre-requisite  is the existence of a well
defined 3-space (i.e. a clock synchronization convention) supporting
the Cauchy data.\medskip

In Galilei space-time there is no problem: time and  Euclidean
3-space are absolute.\medskip

Instead there is no intrinsic notion of 3-space, simultaneity, 1-way
velocity of light (two distant clocks are involved) in the absolute
Minkowski space-time: only the  light-cone is intrinsically given as
the locus of incoming and outgoing radiation. The light postulate
says that the 2-way (only one clock is involved) velocity of light
$c$ is isotropic and constant. Its codified value replaces the rods
(i.e. the standard of length) in modern metrology, where an atomic
clock gives the standard of time. Einstein's $1/2$ synchronization
convention \footnote{An inertial observer A send a ray of light at
$x^o_i$ towards the (in general accelerated) observer B; the ray is
reflected towards A at a point P of B world-line and then reabsorbed
by A at $x^o_f$; by convention P is synchronous with the mid-point
between emission and absorption on A's world-line, i.e. $x^o_P =
x^o_i + {1\over 2}\, (x^o_f - x^o_i)$.} selects the Euclidean
3-spaces $x^o = c\, t = const.$ of the inertial frames centered on
inertial observers: only in this case the 2-way and 1-way light
velocities coincide. However, with realistic accelerated observers
the convention breaks down and till recently there was no definition
of global non-inertial 3-spaces due to the coordinate singularities
present in the {\it 1+3 point of view} (only the world-line of a
time-like observer is given) both with Fermi coordinates (crossing
of the 3-spaces) and rotating frames (the horizon problem of the
rotating disk).
\medskip

In Ref.\cite{1} the theory of global non-inertial frames is fully
developed in the {\it 3+1 point of view}: besides the observer
world-line one gives an admissible 3+1 splitting of Minkowski
space-time, i.e. a nice foliation whose leaves are instantaneous
3-spaces. Lorentz-scalar observer-dependent radar 4-coordinates
$\sigma^A = (\tau; \sigma^r)$ are used: $\tau$ is an arbitrary
increasing function of the observer proper time and $\sigma^r$ are
curvilinear 3-coordinates on the 3-spaces $\Sigma_{\tau}$ with the
observer as origin. Each 3-space is asymptotically Euclidean with
asymptotic inertial observers at spatial infinity. The inverse
transformation $\sigma^A \mapsto x^{\mu} = z^{\mu}(\tau, \sigma^r)$
defines the embeddings of the 3-spaces $\Sigma_{\tau}$ into
Minkowski space-time and the induced 4-metric is $g_{AB}[z(\tau,
\sigma^r)] = [z^{\mu}_A\, \eta_{\mu\nu}\, z^{\nu}_B](\tau,
\sigma^r)$, where $z^{\mu}_A = \partial\, z^{\mu}/\partial\,
\sigma^A$ and ${}^4\eta_{\mu\nu} = \sgn\, (+---)$ is the flat metric
($\sgn = \pm 1$ according to either the particle physics $\sgn = 1$
or the general relativity $\sgn = - 1$ convention). While the
4-vectors $z^{\mu}_r(\tau ,\sigma^u)$ are tangent to
$\Sigma_{\tau}$, so that the unit normal $l^{\mu}(\tau ,\sigma^u)$
is proportional to $\epsilon^{\mu}{}_{\alpha \beta\gamma}\,
[z^{\alpha}_1\, z^{\beta}_2\, z^{\gamma}_3](\tau ,\sigma^u)$, we
have $z^{\mu}_{\tau}(\tau ,\sigma^r) = [N\, l^{\mu} + N^r\,
z^{\mu}_r](\tau ,\sigma^r)$ ($N(\tau ,\sigma^r) = \sgn\,
[z^{\mu}_{\tau}\, l_{\mu}](\tau ,\sigma^r)$ and $N_r(\tau ,\sigma^r)
= - \sgn\, g_{\tau r}(\tau ,\sigma^r)$ are the lapse and shift
functions).\medskip

The foliation is nice and admissible if it satisfies the conditions:
\hfill\break
 1) $N(\tau ,\sigma^r) > 0$ in every point of
$\Sigma_{\tau}$ (the 3-spaces never intersect, avoiding the
coordinate singularity of Fermi coordinates);\hfill\break
 2) $\sgn\, {}^4g_{\tau\tau}(\tau ,\sigma^r) > 0$, so to avoid the
 coordinate singularity of the rotating disk, and with the positive-definite 3-metric
${}^3g_{rs}(\tau ,\sigma^u) = - \sgn\, {}^4g_{rs}(\tau ,\sigma^u)$
having three positive eigenvalues (these are the M$\o$ller
conditions \cite{1});\hfill\break
 3) all the 3-spaces $\Sigma_{\tau}$ must tend to the same space-like
 hyper-plane at spatial infinity (so that there are always asymptotic inertial
observers to be identified with the fixed stars).\medskip

These conditions imply that global {\it rigid} rotations are
forbidden in relativistic theories. In Ref.\cite{1} there is the
expression of the admissible embedding corresponding to a 3+1
splitting of Minkowski space-time with parallel space-like
hyper-planes (not equally spaced due to a linear acceleration)
carrying differentially rotating 3-coordinates without the
coordinate singularity of the rotating disk. It is the first
consistent global non-inertial frame of this type.

\bigskip

As shown in Refs.\cite{1,2} every isolated system (particles,
strings, fluids, fields) admitting a Lagrangian ${\cal L}(matter)$
can be reformulated as a {\it parametrized Minkowski theory}, in
which the new embedding-dependent  Lagrangian is ${\cal L}(matter,
g_{AB}[z])$. This action is invariant under frame-preserving
4-diffeomorphisms, so that the embeddings are {\it gauge variables}
and the ten components of $g_{AB}[z]$ are the special-relativistic
{\it inertial potentials} \footnote{They generate the relativistic
apparent forces in the non-inertial frame and in the
non-relativistic limit they reduce to the Newtonian inertial
potentials. The extrinsic curvature ${}^3K_{rs}(\tau, \sigma^u) =
[{1\over {2\, N}}\, (N_{r|s} + N_{s|r} - \partial_{\tau}\,
{}^3g_{rs})](\tau, \sigma^u)$, describing the {\it shape} of the
instantaneous 3-spaces of the non-inertial frame as embedded
3-manifolds of Minkowski space-time, is a functional of the
independent inertial potentials ${}^4g_{AB}$.}. A change of clock
synchronization (of the shape of $\Sigma_{\tau}$) and/or of the
3-coordinates into the 3-spaces is a gauge transformation: physics
does not change, only the appearances of phenomena change.\medskip

In this formulation the description of matter has to be done with
quantities which know the instantaneous 3-spaces $\Sigma_{\tau}$.
For instance a Klein-Gordon field $\tilde \phi (x)$ will be replaced
with $\phi(\tau ,\sigma^r) = \tilde \phi (z(\tau ,\sigma^r))$; the
same for every other field. Instead for a relativistic particle with
world-line $x^{\mu}(\tau )$ we must make a choice of its energy
sign: then it will be described by 3-coordinates $\eta^r(\tau )$
defined by the intersection of the world-line with $\Sigma_{\tau}$:
$x^{\mu}(\tau ) = z^{\mu}(\tau ,\eta^r(\tau ))$. Differently from
all the previous approaches to relativistic mechanics, the dynamical
configuration variables are the 3-coordinates $\eta^r_i(\tau)$ and
not the world-lines $x^{\mu}_i(\tau)$ (to rebuild them in an
arbitrary frame we need the embedding defining that frame!).

\medskip

The {\it inertial rest-frame instant form} of the isolated system
\cite{1,2} is obtained by restricting the embedding to the inertial
rest-frame centered on the Fokker-Pryce center of inertia: its
Euclidean Wigner-covariant 3-spaces are orthogonal to the conserved
4-momentum of the isolated system. Every isolated system can be
described as a decoupled non-local (and therefore {\it
un-observable}) canonical non-covariant Newton-Wigner external
center of mass \footnote{It is convenient to replace it with its
initial value, namely with the Jacobi data of the Hamilton-Jacobi
formulation.}, with an associated external realization of the
Poincare' algebra, carrying a pole-dipole structure: the invariant
mass $M$ and the rest spin ${\vec {\bar S}}$ of the isolated system.
By construction, they depend upon Wigner-covariant relative
variables describing the internal dynamics of the isolated system
\footnote{Inside the Wigner 3-spaces there is an unfaithful internal
realization of the Poincare' algebra, determined by the
energy-momentum tensor, whose energy is the invariant mass and whose
angular momentum is the rest spin. The internal 3-momentum vanishes
being the rest-frame condition. The internal center of mass inside
the Wigner 3-spaces is eliminated by the vanishing of the internal
(interaction-dependent) Lorentz boosts, avoiding a double counting
of this collective variable.}.\medskip

The world-lines $x^{\mu}_i(\tau)$ of the particles are derived
(interaction-dependent) quantities and in general they do not
satisfy vanishing Poisson brackets: already at the classical level a
non-commutative structure emerges!

\bigskip

The definition of relativistic atomic physics (scalar
positive-energy charged particles plus the electro-magnetic field in
the radiation gauge with Grassmann-valued electric charges to
regularize self-energies) and of its Poincare' generators becomes
possible \cite{3,4,5} in this framework. The identification of the
Darwin potential, to be added to the Coulomb one, in this classical
setting establishes a contact with the theory of relativistic bound
states, whose constituents must be synchronized (absence of relative
times).

\bigskip

Also a new formulation of {\it relativistic quantum mechanics and
entanglement} was given \cite{6}. The use of the static Jacobi data
for the external center of mass avoids the causality problems
connected with the instantaneous spreading of wave packets. Due to
the need of clock synchronization for the definition of the
instantaneous 3-spaces, the Hilbert space $H = H_{com, HJ} \otimes
H_{rel}$ ($H_{com, HJ}$ is the Hilbert space of the external center
of mass in the Hamilton-Jacobi formulation, while $H_{rel}$ is the
Hilbert space of the internal relative variables) is not unitarily
equivalent to $H_1 \otimes H_2 \otimes ...$, where $H_i$ are the
Hilbert spaces of the individual particles. As a consequence, at the
relativistic level the zeroth postulate of non-relativistic quantum
mechanics does not hold: the Hilbert space of composite systems is
not the tensor product of the Hilbert spaces of the sub-systems. The
non validity of the zeroth postulate and the {\it non-locality} of
Poincare' generators imply a {\it kinematical non-locality} and a
{\it kinematical spatial non-separability} introduced by special
relativity, which reduce the relevance of {\it quantum non-locality}
in the study of the foundational problems of quantum mechanics which
have to be rephrased in terms of relative variables.

\bigskip

The replacement of clock synchronization with an admissible 3+1
splitting can be used also in general relativity (GR), where also
the space-time becomes dynamical \cite{7}, being determined by
Einstein equations modulo 4-coordinate transformations (the gauge
group of GR). We will define global  non-inertial frames (the only
ones existing in the large in GR due to the equivalence principle)
with admissible 3+1 splittings and radar 4-coordinates in globally
hyperbolic, asymptotically Minkowskian space-times in the framework
of ADM canonical gravity. With suitable boundary conditions,
eliminating super-translations \cite{8}, the asymptotic symmetries
reduce to the ADM Poincare' group \footnote{For $G = 0$ it reduces
to the Poincare' group of the matter in Minkowski non-inertial
frames. In this way, after a restriction to inertial frames we can
recover all the results of the standard model of elementary
particles, which are connected with properties of the
representations of the Poincare' group in inertial frames of
Minkowski space-time.} and the non-Euclidean 3-spaces are orthogonal
to the conserved ADM 4-momentum at spatial infinity \cite{9}: this
is a {\it non-inertial rest frame} of the 3-universe (see
Ref.\cite{1} for the non-inertial rest-frame instant form in special
relativity). There are asymptotic inertial observers with spatial
axes identified by means of the fixed stars of star catalogues.

\medskip

As a consequence, the 3-universe (the isolated system "gravitational
field plus matter") can be described as a decoupled non-covariant
non-observable external pseudo-particle carrying a pole-dipole
structure, whose mass and spin are  identified by the ADM weak
energy and by the ADM angular momentum. Instead the ADM 3-momentum
vanishes, since this determines the rest-frame condition. The
vanishing of the ADM Lorentz boosts eliminate the internal center of
mass of the 3-universe.\medskip

In absence of matter Christodoulou - Klainermann space-times
\cite{10} are compatible with this description.

\medskip

Now the dynamical variable is not the embedding but the 4-metric,
which determines the dynamical chrono-geometrical structure of
space-time by means of the line element: it teaches  to massless
particles which are the allowed trajectories in each point
\footnote{In 2013 the ESA-ACES mission \cite{11} on the
synchronization of atomic clocks between Earth and the Space Station
will make the first precision measurement of the gravitational
redshift created by the geo-potential, i.e. of the $1/c^2$
modifications of the Minkowski light-cone. Every approach to quantum
gravity will have to reproduce these data. A varying light-cone is a
non-perturbative effect in every quantum field theory, string
included, because to define the Fock space one needs the Fourier
decomposition of fields on a fixed background space-time with a
fixed light-cone. On the other hand in loop quantum gravity one has
still to find a well defined coarse graining identifying Minkowski
space-time and perturbations around it.}. Since tetrad gravity is
more natural for the coupling of gravity to the fermions, the
4-metric is decomposed in terms of cotetrads, ${}^4g_{AB} =
E_A^{(\alpha)}\, {}^4\eta_{(\alpha)(\beta)}\, E^{(\beta)}_B$
\footnote{$(\alpha)$ are flat indices; the cotetrads
$E^{(\alpha)}_A$ are the inverse of the tetrads $E^A_{(\alpha)}$
connected to the world tetrads by $E^{\mu}_{(\alpha)}(x) =
z^{\mu}_A(\tau, \sigma^r)\, E^A_{(\alpha)}(z(\tau, \sigma^r))$.},
and the ADM action, now a functional of the 16 fields
$E^{(\alpha)}_A(\tau, \sigma^r)$, is taken as the action for ADM
tetrad gravity \cite{9}. This leads to an interpretation of gravity
based on a congruence of time-like observers endowed with
orthonormal tetrads: in each point of space-time the time-like axis
is the  unit 4-velocity of the observer, while the spatial axes are
a (gauge) convention for observer's gyroscopes.

\medskip

In canonical ADM tetrad gravity there are 16 fields, 16 conjugate
momenta, 14 first-class constraints, generators of Hamiltonian gauge
transformations, 14 gauge variables, the {\it GR inertial effects}
and 2+2 physical variables, the {\it tidal effects} (the
gravitational waves after linearization). As shown in
Refs.\cite{9,12}, in our family of space-times the Dirac Hamiltonian
turns out to be the {\it weak} ADM energy \footnote{It is a volume
integral over 3-space of a coordinate-dependent energy density. It
is weakly equal to the {\it strong} ADM energy, which is a flux
through a 2-surface at spatial infinity.} plus constraints.
Therefore in this family of space-times there is {\it not a frozen
picture}, like in the family of spatially compact without boundary
space-times considered in loop quantum gravity, where the Dirac
Hamiltonian is a combination of constraints.

\medskip

In Ref.\cite{9}  a York canonical basis, adapted to ten first-class
constraints, was identified: this allows for the first time to get
the explicit identification of the inertial and tidal variables. It
implementes the York map of Ref.\cite{13} and diagonalizes the
York-Lichnerowicz approach \cite{14}. Its final form is
($\alpha_{(a)}(\tau, \sigma^r)$ are angles, ${}^3e_{(a)r}(\tau,
\sigma^r)$ are cotriads on the 3-space, $1 + n(\tau, \sigma^r)$ and
${\bar n}_{(a)}(\tau, \sigma^r)$ are the lapse and shift functions
respectively)

\bea
 &&\begin{minipage}[t]{4 cm}
\begin{tabular}{|ll|ll|l|l|l|} \hline
$\varphi_{(a)}$ & $\alpha_{(a)}$ & $n$ & ${\bar n}_{(a)}$ &
$\theta^r$ & $\tilde \phi$ & $R_{\bar a}$\\ \hline
$\pi_{\varphi_{(a)}} \approx0$ &
 $\pi^{(\alpha)}_{(a)} \approx 0$ & $\pi_n \approx 0$ & $\pi_{{\bar n}_{(a)}} \approx 0$
& $\pi^{(\theta )}_r$ & $\pi_{\tilde \phi} = {{c^3}\over {12\pi G}}\, {}^3K$ & $\Pi_{\bar a}$ \\
\hline
\end{tabular}
\end{minipage}\nonumber \\
 &&{}\nonumber \\
 &&{}\nonumber \\
 &&{}^3e_{(a)r} = R_{(a)(b)}(\alpha_{(c)})\, {}^3{\bar e}_{(b)r} =
 R_{(a)(b)}(\alpha_{(c)})\, V_{rb}(\theta^i)\,
 {\tilde \phi}^{1/3}\, e^{\sum_{\bar a}^{1,2}\, \gamma_{\bar aa}\, R_{\bar a}},\nonumber \\
 &&{}^4g_{\tau\tau} = \sgn\, [(1 + n)^2 - \sum_a\, {\bar n}^2_{(a)}],
 \qquad {}^4g_{\tau r} = - \sgn\, {\bar
 n}_{(a)}\, {}^3{\bar e}_{(a)r},\nonumber \\
 &&{}^4g_{rs} = - \sgn\, {}^3g_{rs} = - \sgn\, {\tilde \phi}^{2/3}\,
 \sum_a\, V_{ra}(\theta^i)\, V_{sa}(\theta^i)\,
 e^{2\, \sum_{\bar a}^{1,2}\, \gamma_{\bar aa}\, R_{\bar
 a}},\nonumber \\
 &&{}\nonumber \\
 \eea

In this York canonical basis the {\it inertial effects} are
described by the arbitrary gauge variables $\alpha_{(a)}$,
$\varphi_{(a)}$, $1 + n$, ${\bar n}_{(a)}$, $\theta^i$, ${}^3K$,
while the {\it tidal effects}, i.e. the physical degrees of freedom
of the gravitational field, by the two canonical pairs $R_{\bar a}$,
$\Pi_{\bar a}$, $\bar a =1,2$. The momenta $\pi_r^{(\theta)}$ and
the 3-volume element $\tilde \phi = \sqrt{det\, {}^3g_{rs}}$ have to
be found as solutions of the super-momentum and super-hamiltonian
(i.e. the Lichmerowicz equation) constraints, respectively.\medskip

The gauge variables $\alpha_{(a)}$, $\varphi_{(a)}$ parametrize the
extra O(3,1) gauge freedom of the tetrads (the gauge freedom for
each observer to choose three gyroscopes as spatial axes and to
choose the law for their transport along the world-line). We have
studied in detail the Schwinger time gauges where we impose the
gauge fixings $\varphi_{(a)}(\tau, \sigma^r) \approx 0$,
$\alpha_{(a)}(\tau, \sigma^r) \approx 0$ so that the tetrads become
adapted to the 3+1 splitting (the time-like tetrad coincides with
the unit normal to the 3-space).

\medskip

The gauge angles $\theta^i$ (i.e. the director cosines of the
tangents to the three coordinate lines in each point of
$\Sigma_{\tau}$) describe the freedom in the choice of the
3-coordinates $\sigma^r$ on each 3-space: their fixation implies the
determination of the shift gauge variables ${\bar n}_{(a)}$, namely
the appearances of gravito-magnetism in the chosen 3-coordinate
system.\medskip

One momentum is a gauge variable (a reflex of the Lorentz
signature): the {\it York time}, i.e. the trace ${}^3K(\tau,
\sigma^r)$ of the {\it extrinsic curvature} of the non-Euclidean
3-spaces as 3-sub-manifolds of space-time. This inertial effect
(absent in Newtonian gravity with its absolute Euclidean 3-space)
describes the GR remnant of the special-relativistic gauge freedom
in clock synchronization. Its fixation determines the lapse
function.\medskip

In the York canonical basis the Hamilton equations generated by the
Dirac Hamiltonian $H_D = {\hat E}_{ADM} + (constraints)$ are divided
in four groups: A) four contracted Bianchi identities, namely the
evolution equations for $\tilde \phi$ and $\pi_i^{(\theta)}$ (they
say that given a solution of the constraints on a Cauchy surface, it
remains a solution also at later times); B) four evolution equation
for the four basic gauge variables $\theta^i$ and ${}^3K$: these
equations determine the lapse and the shift functions once four
gauge fixings for the basic gauge variables are added; C) four
evolution equations for the tidal variables $R_{\bar a}$, $\Pi_{\bar
a}$; D) the Hamilton equations for matter, when present.

\medskip

Once a gauge is completely fixed, the Hamilton equations become
deterministic. Given a solution of the super-momentum and
super-Hamiltonian constraints and the Cauchy data for the tidal
variables on an initial 3-space, we can find a solution of
Einstein's equations in radar 4-coordinates adapted to a time-like
observer. To it there is associated a special 3+1 splitting of
space-time with dynamically selected instantaneous 3-spaces in
accord with Ref.\cite{7}. Then we can get pass to adapted world
4-coordinates ($x^{\mu} = z^{\mu}(\tau, \sigma^r) = x^{\mu}_o +
\epsilon^{\mu}_A\, \sigma^A$) and we can describe the solution in
every 4-coordinate system by means of 4-diffeomorphisms.

\medskip

In Ref.\cite{15} we study the coupling of N charged scalar particles
plus the electro-magnetic field to ADM tetrad gravity  in this class
of asymptotically Minkowskian space-times without
super-translations. To regularize the self-energies both the
electric charge and the sign of the energy of the particles are
Grassmann-valued. The introduction of the non-covariant radiation
gauge allows to reformulate the theory in terms of transverse
electro-magnetic fields and to extract the generalization of the
Coulomb interaction among the particles in the Riemannian
instantaneous 3-spaces of global non-inertial frames.

\medskip

From  the Hamilton equations  in the York canonical basis \cite{15},
followed by a Post-Minkowskian linearization with the asymptotic
flat Minkowski 4-metric at spatial infinity as background, it has
been possible to develop a theory of gravitational waves with
asymptotic background propagating in the non-Euclidean 3-spaces
$\Sigma_{\tau}$ of a family of {\it non-harmonic 3-orthogonal}
gauges \footnote{The 3-metric in $\Sigma_{\tau}$ is diagonal like in
astronomical frames GCRS and BCRS.} parametrized by the values of
the York time ${}^3K(\tau, \sigma^r)$ (the left gauge freedom in the
shape of $\Sigma_{\tau}$).

\bigskip

The conceptual problem of the GR gauge freedom in the choice of the
4-coordinates is {\it solved at the experimental level inside the
Solar system by the choice of a convention for the description of
matter}: a) for satellites near the Earth (like the GPS ones) one
uses NASA 4-coordinates compatible with the terrestrial ITFR2003 and
geocentric GCRS IAU2000 \cite{16} frames; b) for planets  in the
Solar System one uses the barycentric BCRS-IAU2000 \cite{16} frame.
These frames are compatible with "quasi-inertial frames" in
Minkowski space-time. These are metrological choices like the choice
of a certain  atomic clock as standard of time.\medskip

In astronomy the positions of stars and galaxies are determined from
the data (luminosity, light spectrum, angles) on the sky as living
in a 4-dimensional nearly-Galilei space-time with the celestial ICRS
\cite{16} frame  considered as a "quasi-inertial frame" (all
galactic dynamics is Newtonian gravity), in accord with the standard
FRW $\Lambda$CDM cosmological model when the constant intrinsic
3-curvature of 3-spaces is zero (as implied by the CMB
data\cite{17}). To reconcile all the data with this 4-dimensional
reconstruction one must postulate the existence of dark matter and
dark energy as the dominant components of the classical universe
after the recombination 3-surface!

\medskip

Our proposal is to define a {\it Post-Minkowskian ICRS} with
non-Euclidean 3-spaces, whose intrinsic 3-curvature (due essentially
to gravitational waves) is small, in such a way that the York time
be (at least partially) fitted to the observational data implying
the presence of dark matter. As shown in Ref.\cite{15} the
Post-Newtonian limit of the Post-Minkowskian Hamilton equations of
particles in this family of gauges reproduces Kepler equations plus
a $v/c$ term depending on the York time (the arbitrary gauge
function). Therefore there is the concrete possibility (under
investigation) to explain the rotation curves of galaxies \cite{18}
as a {\it relativistic inertial effect inside Einstein GR} (choice
of a York time compatible with observations \cite{19}) without
modifications: a) of Newton gravity like in MOND \cite{20}; b) of GR
like in $f(R)$ theories \cite{21}; c) of particle physics with the
introduction of WIMPS \cite{22}. Then, the next step will be to
study the dependence on the York time of quantities like redshift,
luminosity distance, gravitational lensing.... and to see which
information on the York time can be extracted from the data
supporting dark energy.
\medskip

In conclusion the reformulation of clock synchronization as the
existence of well-defined non-Euclidean 3-spaces with the gauge
freedom of the York time plus the proposed  way out from the GR
gauge problem using the observational metrological conventions may
help in reducing the dark side of the universe to a relativistic
inertial effect inside Einstein GR by means of a Post-Minkowskian
definition of ICRS, which will be also useful for the ESA-GAIA
mission \cite{23} (cartography of the Milky Way) and for the
possible anomalies inside the Solar System \cite{24}.
\medskip

Finally the transition to cosmology should be done with approaches
of the type of backreaction \cite{25}.

\end{document}